\documentclass[aps, prl,10pt,twocolumn,superscriptaddress]{revtex4-1}

\usepackage{graphicx}
\usepackage{tabularx}
\usepackage{color}
\usepackage{amsmath}
\usepackage{comment}
\newcommand{\be}{\begin{equation}}
\newcommand{\ee}{\end{equation}}
\newcommand{\bea}{\begin{eqnarray}}
\newcommand{\eea}{\end{eqnarray}}
\newcommand{\ep}{\epsilon}
\usepackage{dcolumn}
\usepackage{hyperref}
\usepackage{bm}
\usepackage{epsf}
\usepackage{subfigure}
\usepackage{braket}
\usepackage{amsfonts}

\begin{document}
\title{ Bounds on fluctuations for finite-time quantum Otto cycle} 
\author{Sushant Saryal}
\affiliation{Department of Physics, Indian Institute of Science Education and Research Pune, Dr. Homi Bhabha Road, Ward No. 8, NCL Colony, Pashan, Pune, Maharashtra 411008, India}
\author{Bijay Kumar Agarwalla}
\email{bijay@iiserpune.ac.in}
\affiliation{Department of Physics, Indian Institute of Science Education and Research Pune, Dr. Homi Bhabha Road, Ward No. 8, NCL Colony, Pashan, Pune, Maharashtra 411008, India}

\date{\today}
%====================================
\begin{abstract}
%====================================
For finite-time quantum Otto heat engine with working fluid consisting of either a (i) qubit or (ii) a harmonic oscillator, we show that the relative fluctuation of output work is always greater than the corresponding relative fluctuation of input heat absorbed from the hot bath. As a result, the ratio between the work fluctuation and the input heat fluctuation receives a lower bound in terms of the square value of the average efficiency of the engine. The saturation of the lower bound is received in the quasi-static limit of the engine and can be shown for a class of working fluids that follow a scale-invariant energy eigenspectra under driving.  
\end{abstract}
\maketitle

\section{Introduction}  The quest to build most efficient and powerful heat engine led Sadi Carnot\cite{Carnot} to pioneer the subject what is known today as  {\it Thermodynamics} \cite{Carnot-1, Carnot-2}. Although initial development of the subject were motivated by engineering optimization problem, thermodynamics remained as one of the fundamental physical theory in science. In fact its core principles have survived both relativity and quantum revolution. One of the central result of thermodynamics is that efficiency of any engine operating between hot and cold  reservoirs with temperatures $T_h$ and $T_c$, respectively, is upper bounded by Carnot efficiency, $\eta_C = 1- T_c/T_h$. Traditionally thermodynamics was only concerned with average quantities as fluctuation can be ignored for large systems ,  for example, steam engines, automobile engines etc. But with the rapid technological development of miniturization of devices and advancement in accessing very low temperatures one  can no longer ignore fluctuations of thermal and/or quantum origin \cite{eff-1,scovil,Uzdin-theory,engine-expt1,engine-expt-spin,engine-expt-spin-osc,engine-refg,Eric-engine}.  Over the  last three decades due to the discovery of fluctuation theorems \cite{st-thermo1,st-thermo2,Q-thermo2, fluc-1,fluc-2, fluc-3}  we have taken a big leap in understanding fluctuations of very large class of systems driven out-of-equilibrium. 

Very recently, for out-of-equilibrium systems, {\it thermodynamic uncertainty relations} (TURs) \cite{Barato:2015:UncRel, trade-off-engine, Gingrich:2016:TUP,Falasco,Garrahan18,Timpanaro,Saito-TUR,Junjie-TUR, Agarwalla-TUR} provided lower bound on the relative fluctuations of integrated currents (heat, particle, energy etc.) in terms of the net entropy production. In other words, TUR restricts optimization of relative fluctuations and entropy production in an arbitrary manner by providing a trade-off relation between these quantities.  As a consequence of this result, a continuous heat engine operating in a non-equilibrium steady state follows a trade-off relation involving its efficiency, output power and power fluctuations \cite{trade-off-engine}. For a similar setup operating as an engine, it was recently shown by some of us that, in the linear response regime, relative fluctuation of work current is always lower bounded by the input heat current \cite{Universal}.  In this paper, we consider a finite-time quantum Otto engine setup and show that a bound similar to Ref.~(\cite{Universal}) exists. 
In particular, we show, for two prototypical systems driven arbitrarily, that, the ratio of work fluctuation and input heat fluctuation from the hot bath receives a lower bound which is determined by the square of the average efficiency of the engine. The equality of the bound is received in the quasi-static (QS) limit and can be shown for a class of working fluids following a scale invariant energy eigenspectra.

The plan of the paper is as follows: We first introduce the quantum Otto cycle along with the projective measurement scheme that allows us to construct the probability distribution function to study fluctuations. Next in the QS limit, we derive a general joint cumulant generating function of heat and work for scale-invariant  driven hamiltonian and show that the ratio of $n-$th cumulant of output work and $n-$th cumulant of input heat is exactly equal to the $n$-th power of the average efficiency and consequently upper bounded by the $n$-th power of the Carnot efficiency. Next we provide two paradigmatic examples (a two-level system (TLS) and a harmonic oscillator (HO)) of non-adiabatic driving of quantum Otto cycle and show that relative fluctuations of work are always lower bounded by relative fluctuations of heat whenever the Otto cycle works as engine. Finally we summarise our central results. We delegate certain technical details to the appendix.

\section{Universal Quantum Otto Cycle}
\begin{figure}[h]
\includegraphics[width=\columnwidth]{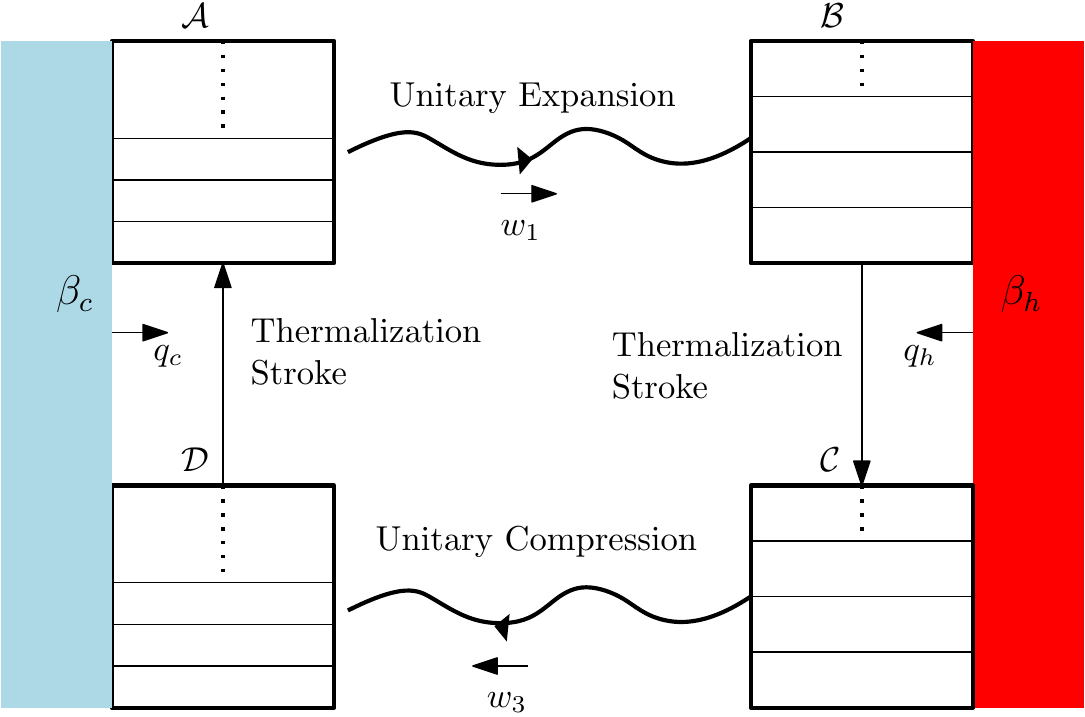}
\caption{(Color online): Schematic of a four-stroke Quantum Otto cycle. For a detailed description about the cycle please refer to the text.  As per our convention, energy flowing towards the working fluid is considered as positive. The cycle operates as a heat engine when  $\langle w\rangle = \langle w_1\rangle + \langle w_3\rangle <0$ and $\langle q_h\rangle > 0$.}
\label{otto-scheme}
\end{figure}

We consider a standard four-stroke quantum Otto cycle \cite{o1,o2,o3,o4,o5,o6,o7,o8,o9,o10,Lutz1}, as illustrated in Fig.~(\ref{otto-scheme}).  The working fluid is initially ($t=0)$ thermalized  by placing it in a weak contact with a cold reservoir at inverse temperature $\beta_c= 1/T_c$ ($k_B$ is set to unity). The fluid is then separated from the bath and is subjected to four strokes. (i) 
{\it Unitary expansion stroke} ($A \to B$) -- 
In this stroke, the working fluid expands unitarily under a time-dependent driving that takes the initial hamiltonian $H_0$ to a final hamiltonian $H_{\tau}$ in a time duration $\tau$. The working fluid, in this step, consumes an amount of work $w_1$ which is not a fixed number but rather a stochastic quantity due to the random thermal initial condition and possible quantum fluctuations during the unitary evolution. 
(ii) {\it Isochoric heating stroke} ($B \to C$)--
During this step, the working medium is put in weak contact with a hot bath at inverse temperature $\beta_h$ to achieve full thermalization.  The Hamiltonian for the working fluid therefore remains the same while the fluid absorbs an amount of  heat $q_h$. Here,  we assume that the interaction time with the bath is long enough to achieve equilibration.
(iii) {\it Unitary compression stroke} ($C \to D$) -- In the next stroke, the system is detached from the hot bath and unitarily compressed via driving the working fluid back to the initial hamiltonian $H_0$ starting from $H_{\tau}$ while the fluid consuming an amount of work $w_3$.  For simplicity, we assume that the time duration for this stroke is the same as the expansion stroke.  In this study, we are going to assume that the compression protocol is a time-reversed version of the corresponding expansion protocol.
(iv) {\it Isochoric cooling stroke } ($D \to A$) -- In the final stroke, the fluid is put in contact with a cold bath at inverse temperature $\beta_c$ to reach equilibrium and thereby closing the cycle. 
It is important to note that,  as per our convention (see Fig.~(\ref{otto-scheme})),  energy flowing into the fluid is always considered to be positive. From here onwards, we denote $w=w_1+w_3$ as the net work performed on the working fluid. 

{\it Joint probability distribution for work and input heat in quantum Otto engine--.}  In the quantum regime, a thermodynamically consistent way of studying fluctuations for non-equilibrium systems is via the two-point projective measurement scheme. Such a measurement scheme is also consistent with the quantum fluctuation relations  \cite{st-thermo1,st-thermo2,Q-thermo2, fluc-1,fluc-2, fluc-3}. In fact, very recently, following this scheme, an expression for efficiency statistics for Otto cycle with arbitrary working fluid was obtained \cite{Lutz1}. We follow a similar procedure here. Since in this work we are interested  only in the heat engine regime,  we construct the joint probability distribution $p(w_1,q_h,w_3)$ by performing projective measurements of the respective hamiltonians involving in the first three strokes $(A \to B \to C \to D)$.
We then receive,
\bea
&& p(w_1,q_h,w_3)=\sum_{n mk l} \delta \Big(w_1-(\ep^{\tau}_m-\ep^0_n)\Big) \,\, \delta \Big(q_h-(\ep^{\tau}_k-\ep^{\tau}_m)\Big)  \nonumber \\
 \times&&\delta \Big(w_3-(\ep_l^0-\ep^{\tau}_k)\Big) \, T^{\tau}_{n\to m}\,\,T^{\tau}_{k\to l} \frac{e^{-\beta_c \ep_n^0}}{{\cal Z}_0} \, \frac{e^{-\beta_h \ep_k^{\tau}}}{{\cal Z}_{\tau}}
 \label{prob-dist}
\eea
where $\ep^0_n$ ( $\ep^{\tau}_n$) are the energy eigenvalues of initial (final) hamiltonian during the unitary expansion stroke $A \to B$. Here ${\cal Z}_0= \sum_n \exp(-\beta_c \ep^0_n))$ and ${\cal Z}_{\tau}=\sum_n \exp(-\beta_h \ep^{\tau}_n))$ are the partition functions.  $T^{\tau}_{n\to m}=| \langle m_{\tau} | U_{\rm exp} | n_0 \rangle |^2 $ ($T^{\tau}_{k\to l} =| \langle l_{0} | U_{\rm com} | k_{\tau} \rangle |^2 $) is the transition probability between the eigenstates of $H_0$ and $H_{\tau}$ during the unitary expansion (compression) stoke.
From the above distribution function, the joint distribution for net work $w = w_1+ w_3$ and input heat $q_h$ can also be obtained easily. 
As mentioned before, we focus in the engine regime, (i.e., as per our convention, $\langle w\rangle <0$ and $\langle q_h\rangle > 0$) and correspondingly investigate the bound for the ratio for the output work fluctuation to the input heat fluctuation by defining our central quantity 
\bea
\eta^{(2)}= \frac{\langle w^2 \rangle_c}{\langle q_h^2 \rangle_c }.
\eea
It is important to note that, this definition for the ratio of fluctuations or $\eta^{(2)}$ is different than what follows from the stochastic efficiency definition $\tilde{\eta}^{2}= \langle\frac{ w^2}{q_h^2}\rangle_c$  which was recently investigated in \cite{Lutz1}.
In what follows, we first present universal result for $\eta^{(2)}$ for quasi-static Otto cycle with working fluid satisfying a scaling relation and then extend our study to  the non-adiabatic regime for two paradigmatic models, consisting of (i) a two-level system (TLS) and (ii) a harmonic oscillator (HO). 
\section{ Results}
\subsection{Result I -- Quasi-static limit:} 
Before discussing the most-general situation, we first focus on the quasi-static (QS) driving limit for the unitary strokes for an Otto cycle. In this limit,
one receives universal results for $\eta^{(2)}$.  As per the {\it quantum adiabatic theorem} in the slow-driving limit, the occupation probabilities between the instantaneous energy eignestates do not change with time which imply for the  transition probabilities in Eq.~(\ref{prob-dist}) $ p_{n\to m}=\delta_{nm}$ and $p_{k\to l} = \delta_{kl}$. As a result, the joint distribution of input heat $(q_h)$ and the net work $w=w_1+w_3$ simplifies to,
\bea
p_{\rm QS}(w,q_h)&&=\sum_{n,k} \delta(w-[(\ep^{\tau}_n-\ep^0_n)+(\ep^0_k-\ep^{\tau}_k)]) \nonumber \\
\times&& \delta(q_h - (\ep^{\tau}_k-\ep^{\tau}_n))\, \frac{e^{-\beta_c \ep_n^0}}{{\cal Z}_0} \, \frac{e^{-\beta_h \ep_k^{\tau}}}{{\cal Z}_{\tau}}.
\eea
Instead of looking at a most general eigen-spectra for the driving Hamiltonians, we consider a scale-invariant energy eigenspectra under driving, given as $\ep^{\tau}_n = \ep^{0}_n/\lambda^2_{\tau}$ where $\lambda_{\tau}$ is the scaling factor. Such a scaling can be realized for driving Hamiltonians of the form  $H_{t} = {\bf p}^2/2m + V({\bf x}, \lambda_t)$ with the interaction following a scaling property $V({\bf x}, \lambda_t) = V_0({\bf x}/\lambda_t)/\lambda_t^2$.  Such Hamiltonians represent a broad class of single particle and many-body systems \cite{si1,si2,si3,si4,si5}.  
 Under these conditions, the corresponding characteristic function (CF) $\chi_{\rm QS}(\alpha_1,\alpha_2)$ with $\alpha_1$ and  $\alpha_2$ being the counting parameters for $w$ and $q_h$, simplifies to  
\bea
\chi_{\rm QS}(\alpha_1,\alpha_2)\! =\! \sum_{n,k} e^{i (\ep_n^0 - \ep_k^0) \big[ \frac{1}{\lambda_{\tau}^2} (\alpha_1 - \alpha_2) - \alpha_1 \big]} \, \frac{e^{-\beta_c \ep_n^0}}{{\cal Z}_0} \, \frac{e^{-\beta_h \ep_k^{\tau}}}{{\cal Z}_{\tau}}.\nonumber \\
\eea
A relation between work and heat cumulants immediately follows from it, 
\bea
\langle w^n \rangle_c = (-1)^n \big( 1- \lambda_{\tau}^2 \big)^n \langle q_h^n \rangle_c 
\eea
Consequently the $n$-th order ratio for net work and input heat from the hot bath  is given as,
\bea
\eta^{(n)}_{\rm QS}= (-1)^n \frac{\langle w^n \rangle_c}{\langle q_h^n \rangle_c }=\big(1\!- \lambda^2_{\tau} \big)^n = \langle \eta \rangle_{\rm QS}^n \leq \eta_C^n.
\eea
where $\langle \eta \rangle = \langle -w \rangle / \langle q_h \rangle$ is the standard thermodynamic efficiency which for an Otto engine in the QS limit reduces to $\langle \eta \rangle_{\rm QS} = (1- \lambda_{\tau}^2)$.
Note that the upper bound can be simply obtained by demanding the positivity of the net entropy production for the Otto cycle. 
This is our first central result. A similar exercise can be carried out in the refrigerator regime as well, following the strokes $(C \to D \to A \to B)$  and the corresponding $n-$th order ratio for input heat from cold bath and net work is given as,
 \be
\varepsilon^{(n)}  = \frac{\langle q_c^n \rangle_c}{\langle w^n \rangle_c }= \Big(\frac{\lambda_{\tau}^2}{1- \lambda_{\tau}^2}\Big)^n  =\langle \varepsilon\rangle^{n}_{QS} \leq \Big( \frac{1-\eta_C}{\eta_C}\Big)^n.
\ee
where $\varepsilon^{(1)} = \langle q_c \rangle / \langle w \rangle$ is the average coefficient of performance of an Otto refrigerator which in the QS limit reduces to $\langle \varepsilon\rangle_{QS}= \Big(\frac{\lambda_{\tau}^2}{1- \lambda_{\tau}^2}\Big) $.
It is interesting to note that, an universal {\it upper} bound on $\eta^{(2)}$ was studied in Ref.(\cite{ub1}) for four stroke heat engine (working between two heat baths) with classical working substance. 

\subsection{Result II: Beyond quasi-static limit- non-adiabatic driving} 
Beyond the quasi-static limit, it is non-trivial to derive universal bound for arbitrary working fluid. We therefore focus on two paradigmatic model systems to understand the non-adiabatic driving situation. We consider the working fluid for Otto-engine consisting of  i) a two-level system (TLS), and (ii) a simple harmonic oscillator (HO). Interestingly, a TLS as a working fluids was recently implemented and studied from the perspective of an Otto heat engine \cite{engine-expt-spin}. 

\subsection{(a) Working fluid consisting of a  two-level system}
We first consider a TLS with Hamiltonian evolving unitarily from  $H_A= \frac{1}{2} \omega_0 \sigma_x$ to $H_B= \frac{1}{2} \omega_{\tau} \sigma_y$ during the expansion $(A \to B)$ stroke and back to $H_A$ during the compression stroke $(B \to A)$ ($\hbar$ is set to unity). For the compression stroke we consider here the reverse protocol of the expansion stroke. Here $\sigma_{x,y,z}$ are the standard Pauli matrices, $\omega_{0, \tau}$ denote the angular frequencies with  $\omega_{\tau} > \omega_0$ corresponding to the energy gap expansion.   The evolution of the density operator during the expansion (compression) protocol is governed by a unitary operator $U_{\rm exp} (U_{\rm com} = U^{\dag}_{\rm exp})$. It is not necessary for our case to specify this operator explicitly, as the below results are valid for arbitrary time-dependent protocol under the above mentioned initial and final TLS Hamiltonians.

One can obtain the the joint characteristic function (CF) for the net work $w$ and $q_h$ (please see appendix A) and compute the first and second order cumulants of heat and work. 
We present here the expressions,
\begin{widetext}
\bea
\langle w\rangle&\equiv& \langle w_1 \rangle + \langle w_3 \rangle =  \Big[\frac{\omega_0}{2}+\frac{\omega_{\tau}}{2}(1-2u)\Big] \tanh\Big(\frac{\beta_c \omega_0}{2}\Big) + \Big[\frac{\omega_{\tau}}{2}+\frac{\omega_0}{2} (1-2u)\Big] \tanh \Big(\frac{\beta_h \omega_{\tau}}{2}\Big), \\
\langle q_h \rangle &=&- \frac{\omega_{\tau}}{2} \Big[\tanh\Big(\frac{\beta_h  \omega_{\tau}}{2}\Big)+\tanh\Big(\frac{\beta_c \omega_0}{2}\Big)(1-2u)\Big], \\
\langle w^2 \rangle_c &=&  \frac{1}{2} (\omega_{\tau}+\omega_0)^2-2 \,u \, \omega_{\tau} \omega_0  -\langle w_1 \rangle^2-\langle w_3 \rangle^2,\\
\langle q_h^2 \rangle_c &=& \frac{\omega_{\tau}^2}{4} \Big[2- \tanh^{2}\Big(\frac{\beta_h \omega_{\tau}}{2}\Big) -(1-2u)^2 \tanh^{2}\Big(\frac{\beta_c \omega_0}{2}\Big)\Big],
\label{TLS-eqs}
\eea
\end{widetext}
where $u$ represents the probability of no transition between the final and the initial eigenstates during the unitary strokes. The quasi-static (QS) limit therefore corresponds to $u=1$. It is easy to check from the above expressions that, in the QS limit, one receives $\eta_{\rm QS}^{(2)}= \langle \eta \rangle_{\rm QS}^2 = (1- \omega_0/\omega_{\tau})^2$ which matches with the result obtained in the previous section with a proper scaling factor $\lambda_{\tau}^2= \omega_0/\omega_{\tau}$.

Beyond the QS limit, we provide a rigorous proof in appendix B that while the TLS medium working as a heat engine i.e., under the conditions $\langle w \rangle <0$ and $\langle q_h \rangle >0$, the following quantity 
\bea
\mathcal{A} \equiv \langle w^2 \rangle_c \langle q_h \rangle^2 - \langle q_h^2\rangle_c \langle w \rangle^2 \geq 0
\label{proof-TLS}
\eea 
is always non-negative with the equality sign achieved in the QS limit ($u=1$).
Therefore, for TLS Otto cycle, while operating as an engine, we receive  
\bea
\eta^{(2)} \equiv \frac{\langle w^2 \rangle_c}{\langle q^2_h \rangle_c} \ge \frac{\langle w \rangle^2}{\langle q_h \rangle^2}=\langle \eta \rangle^2.
\label{eta-2-TLS}
\eea
In other words, in the engine regime, $\eta^{(2)}$ is always lower bounded by the square value of the average efficiency.  Another way to interpret Eq.~(\ref{proof-TLS}) is that the relative fluctuation of output work is always greater than relative fluctuations of input heat in the engine regime i.e.,
\bea
\frac{\langle w^2 \rangle_c}{\langle w \rangle^2} \ge \frac{\langle q^2_h \rangle_c}{\langle q_h \rangle^2}.
\eea
At this point, it is important to make a connection with the TUR studies \cite{Barato:2015:UncRel, trade-off-engine, Gingrich:2016:TUP,Falasco,Garrahan18,Timpanaro,Saito-TUR,Junjie-TUR, Agarwalla-TUR} which provide independent bounds on relative fluctuations of individual observables (work, heat) in terms of total entropy production. Here we show that, in the engine regime, these bounds are not independent but rather follows the above relation. 
\begin{figure}
\includegraphics[width=\columnwidth]{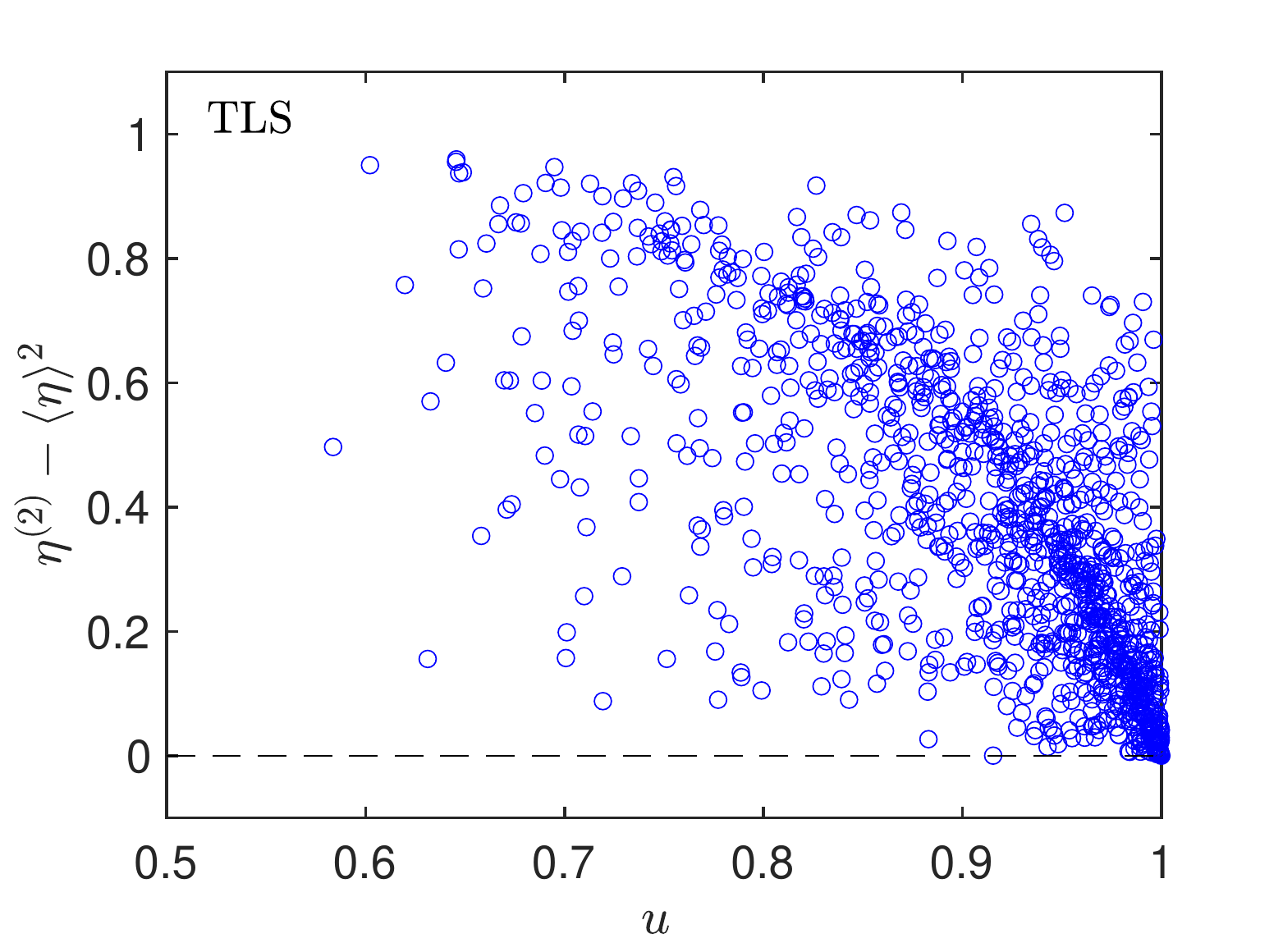}
\caption{Scatter plot of $\eta^{(2)} -  \langle \eta \rangle^2$ for the TLS Otto cycle while operating as a heat engine. All parameters except $u$ here are chosen randomly in the interval between [0, 10]. $u$ is varied randomly between $[0, 1]$. Simulation is done for one million random points.}
\label{otto-2level}
\end{figure}
Note that, we didn't find such a result or proof in the refrigerator regime. This is the second central result of this paper. In Fig.~(\ref{otto-2level}) we present a scatter plot for $\eta^{(2)} -\langle \eta \rangle^2$ for the TLS, while operating as a heat engine, by choosing all the parameters randomly.  It is clear that the lower bound is always respected for this model with the difference disappearing at $u=1$, as expected from Eq.~(\ref{eta-2-TLS}).

\subsection{(b) Working fluid consisting of a harmonic oscillator (HO)}
We next consider another paradigmatic example with working fluid consisting of a single harmonic oscillator. The time-dependent Hamiltonian for the unitary strokes is given as $H(t) = p^2/2m + \frac{1}{2} m \omega^2(t) x^2$ where in this case, the trapping frequency $\omega(t)$ is modulated as a function of time from $\omega_0$ at $t=0$ to $\omega_{\tau} $ at $t=\tau$ during the stroke $A \to B$. For the unitary compression stroke $C \to D$ a reverse protocol is considered which can be obtained from the expansion stroke by replacing $t$ by $\tau -t$. 

\begin{figure}
\includegraphics[width=\columnwidth]{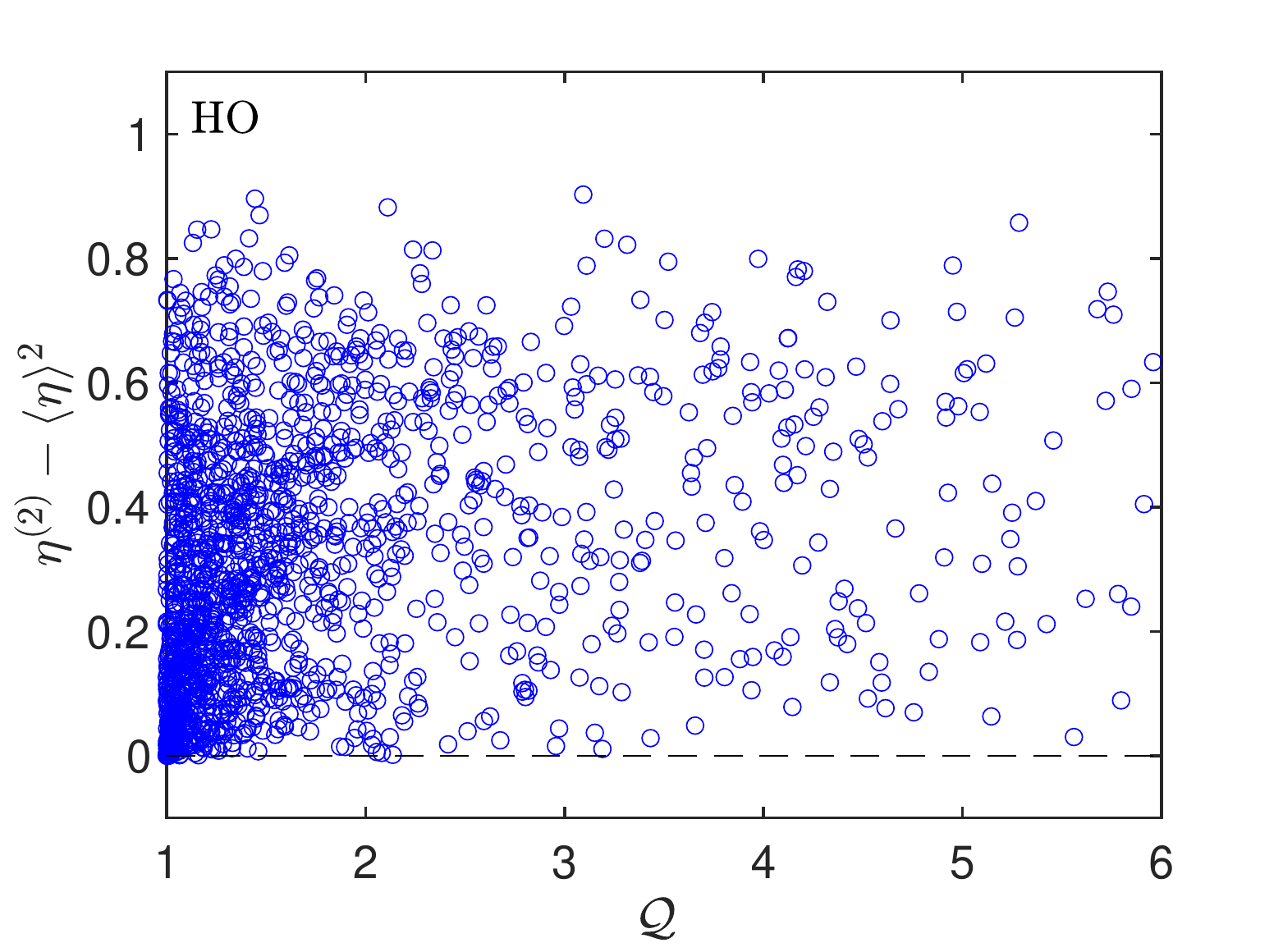}
\caption{Scatter plot of $\eta^{(2)} -  \langle \eta \rangle^2$ for the harmonic oscillator (HO) Otto cycle while operating as a heat engine. All parameters except ${\cal Q}$  are chosen randomly in the interval between [0, 10]. ${\cal Q}$ is varied randomly between $[1,6]$. Simulation is done for one million random points.}
\label{otto-HO}
\end{figure}

The CF for this case can be obtained exactly (please see appendix C). We write down the expressions for the average and the noise for both absorbed heat and net work in the non-adiabatic limit \cite{Denzler-thesis,Deffner-work},
\begin{widetext}
\bea
\langle w\rangle&=& \langle w_1 \rangle + \langle w_3 \rangle =  \frac{1}{2} \Big[ \big({\cal Q} \, \omega_{\tau}-\omega_{0}\big)  \coth \Big(\frac{\beta_c \omega_{0}}{2}\Big) + \big({\cal Q} \, \omega_0-\omega_{\tau}\big) \coth\Big(\frac{\beta_h \omega_{\tau}}{2}\Big) \Big],\\
\langle q_h \rangle &=& \frac{\omega_{\tau}}{2} \Big[\coth\Big(\frac{\beta_h  \omega_{\tau}}{2}\Big)- {\cal Q} \, \coth\Big(\frac{\beta_c \omega_0}{2}\Big)\Big],\\
\langle w^2 \rangle_c &=&   \langle w_1 \rangle^2 + \langle w_3 \rangle^2 - \frac{1}{2} (\omega_0-\omega_{\tau})^2 +( {\cal Q}-1 ) \, \omega_0 \omega_{\tau}  + \frac{1}{4} ({\cal Q}^2 -1) \, \Big[ \omega_{\tau}^2  \coth^{2}\Big(\frac{\beta_c \omega_{0}}{2}\Big) + \omega_{0}^2 \, \coth^{2}\Big(\frac{\beta_h \omega_{\tau}}{2}\Big)\Big],\\
\langle q_h^2 \rangle_c &=& - \frac{\omega_{\tau}^2}{4} \Big[2- \coth^{2}\Big(\frac{\beta_h \omega_{\tau}}{2}\Big) - (2 \, {\cal Q}^2 -1)\, \coth^{2}\Big(\frac{\beta_c \omega_0}{2}\Big)\Big].
\eea
\end{widetext}
The above expressions are valid for arbitrary protocol of $\omega(t)$. Here ${\cal Q} \in [1, \infty]$ is the so-called adiabaticity parameter which characterizes the degree of adiabaticity. The QS limit corresponds to ${\cal Q}=1$ and it is easy to check that $\eta^{(2)}$ saturates the lower bound i.e.,  $\eta_{\rm QS}^{(2)} = \langle \eta \rangle_{\rm QS}^2 = (1- \omega_0/\omega_{\tau})^2$ which is expected as the energy eigenspectra follow the scaling relation. We notice that, for this model as well, in the engine regime, the lower bound is always respected.  In Fig.~(\ref{otto-HO}) we present a scatter plot for $\eta^{(2)} -\langle \eta \rangle^2$ for the HO working fluid in the engine regime by choosing the parameters randomly. It is clear that the lower bound is always respected for this model with the difference disappearing in the QS limit i.e., for ${\cal Q}=1$.

\section{Summary}
In summary, we have investigated bounds on the ratio of non-equilibrium fluctuation for output work and input heat for a finite-time Otto cycle operating as an engine.  We provide a universal result for the ratio $\eta^{(n)}$ in the quasi-static limit which is exactly equal to the $n$-th power of the corresponding average efficiency $\langle \eta \rangle$. In the non-adiabatic limit we show for two paradigmatic models that $\eta^{(n)}$ gets a lower bound determined by $\langle \eta \rangle^2$.  Importantly, this result further connects to the TUR study where as a consequence of the lower bound, the  relative fluctuation of work always surpass the corresponding relative fluctuation of heat absorbed from the hot bath. Future work will be directed towards providing a general proof for the lower bound for arbitrary working fluid operating as an engine in the non-adiabatic regime.

%===================================================================================================
BKA acknowledges the MATRICS grant MTR/2020/000472 from SERB, Government of India. BKA thank the Shastri Indo-Canadian Institute for providing financial support for this research work in the form of a Shastri Institutional Collaborative Research Grant (SICRG). SS acknowledge support from the Council of Scientific \& Industrial Research (CSIR), India (Grant Number 1061651988).  
%=======================================

%==========================================================================================

\vspace{5mm}
\renewcommand{\theequation}{A\arabic{equation}}
\renewcommand{\thefigure}{A\arabic{figure}}
\renewcommand{\thesection}{A\arabic{section}}
\setcounter{equation}{0}  % reset counter

 %\newpage
\appendix 

\section*{Appendix A: Characteristic function for two-level system}
In the main text, we provide the expressions for the first and second order cumulants for the net work and absorbed heat from the hot bath. Here we write down the corresponding characteristic function $\chi_{\rm TLS}(\alpha_1,\alpha_2)$ where $\alpha_1$ and  $\alpha_2$ are  the counting parameters for $w$ and $q_h$, respectively.  We write \cite{Denzler-thesis}, 
\begin{widetext}
\bea
\chi_{\rm TLS} (\alpha_1,\alpha_2)&=&\int_{-\infty}^{\infty} \int_{-\infty}^{\infty} dw \, dq_2\, \,e^{i\alpha_1 w}\, \,e^{i \alpha_2 q_2}\,\, p(w,q_2) \nonumber \\
&=& \frac{1}{{\cal Z}_0 {\cal Z}_{\tau}} \, \Big[2 u^2 \cosh(x_0+x_{\tau})+2 v^2 \cosh(x_0-x_{\tau}) + 2 \, u\, v \cosh(x_{\tau}) \big(e^{-2i\alpha_1 y_0} e^{-x_0}+ e^{2i\alpha_1 y_0} e^{x_0}\big)\nonumber \\
&+& u^2 \big(e^{-x_0+x_{\tau}} e^{-2i\alpha_1 (y_0-y_{\tau})} e^{-2i\alpha_2 y_{\tau}}+ e^{x_0-x_{\tau}} e^{2i\alpha_1(y_0-y_{\tau})} e^{2i\alpha_2 y_{\tau}}\big) \nonumber \\
&+& v^2 \big(e^{-x_0-x_{\tau}} e^{-2i\alpha_1 (y_0+y_{\tau})} e^{2i\alpha_2 y_{\tau}} +  e^{x_0+x_{\tau}} e^{2i\alpha (y_0+y_{\tau})} e^{-2i\alpha_2 y_{\tau}} \Big) \nonumber \\
&+& 2\, u\, v \cosh(x_0)\big( e^{2i(\alpha_1-\alpha_2)y_{\tau}} e^{x_{\tau}} + e^{-2i(\alpha_1-\alpha_2)y_{\tau}} e^{-x_{\tau}} \big) \Big],
\eea 
\end{widetext}
where we denote $x_0=\beta_c\, \omega_0$, $x_{\tau}=\beta_h\, \omega_{\tau}$, and $v=1-u$. All the moments can be obtained from the CF by taking partial derivatives with respect to $\alpha_1, \alpha_2$ i.e., 
\be
\langle w^n q_2^m \rangle = \frac{\partial^{n} \partial^{m}}{\partial(i \alpha_1)^n \, \partial(i \alpha_2)^m}  \chi_{\rm TLS} (\alpha_1,\alpha_2)|_{\alpha_1=\alpha_2=0}.
\ee
In the quasi-static (QS) limit, i.e., for $u=1$, the above CF simplifies to, 
\begin{eqnarray}
\chi^{QS}_{\rm TLS}(\alpha_1,\alpha_2)&&= 1\!+ f_{\tau} \, (1\!-\!f_0)\, \big(e^{-i\alpha_1(\omega_{\tau}-\omega_0)}\, e^{i \alpha_2\omega_{\tau}} \!-\!1\big) \nonumber \\
&& +\!f_0 \, (1\!-\!f_{\tau})\, \big(e^{i\alpha_1(\omega_{\tau}-\omega_0)}\, e^{-i \alpha_2 \omega_{\tau}}\!-\!1\big),
%\chi_{QS}^q(\alpha)&&= 1+ f_{\tau}(1-f_{0})(e^{i \alpha \omega_{\tau}}-1)  +f_0(1-f_{\tau})(e^{-i\alpha \omega_{\tau}}-1)
\end{eqnarray}
%\end{widetext}
where $f_{0}= (e^{\beta_c \omega_{0}}+1)^{-1} (f_{\tau}= (e^{\beta_h \omega_{\tau}}+1)^{-1})$ is the Fermi distribution function for the qubit. 
%As expected, the QS limit corresponds to the saturation of the lower bound for $\eta^{(n)}$  i.e., $\eta^{(n)} = \langle \eta \rangle^n =(1- \omega_0/\omega_{\tau})^n$. As mentioned earlier, this simply follows from the symmetry  $\chi^w(\alpha) = \chi^q(-\alpha)$.
\begin{comment}
 We receive the following expressions in the adiabatic limit,
\begin{widetext}
\bea
\langle w\rangle&=&  \frac{1}{2} \Big[\omega_0-\omega_{\tau}\Big] \Big[\tanh\Big(\frac{\beta_c \omega_0}{2}\Big) - \tanh \Big(\frac{\beta_h \omega_{\tau}}{2}\Big) \Big]\\
\langle q_h \rangle &=&- \frac{\omega_{\tau}}{2} \Big[\tanh\Big(\frac{\beta_h  \omega_{\tau}}{2}\Big)-\tanh\Big(\frac{\beta_c \omega_0}{2}\Big)\Big] \\
\langle w^2 \rangle_c &=&  \frac{1}{4} (\omega_0-\omega_{\tau})^2 \Big[2 - \tanh^{2}\Big(\frac{\beta_h \omega_{\tau}}{2}\Big) -\tanh^{2}\Big(\frac{\beta_c \omega_0}{2}\Big)\Big]\\
\langle q_h^2 \rangle_c &=& \frac{\omega_{\tau}^2}{4} \Big[2- \tanh^{2}\Big(\frac{\beta_h \omega_{\tau}}{2}\Big) -\tanh^{2}\Big(\frac{\beta_c \omega_0}{2}\Big)\Big]
\eea
\end{widetext}
\end{comment}

%where $y_0=\frac{\sqrt{4\lambda(0)^2+\omega^2}}{2}$ and $y_{\tau}=\frac{\sqrt{4\lambda(0)^2+\omega^2}}{2}$ are half of energy gap of initial and final hamiltonian(after driving time $\tau$) repectively. $x_1=\beta_1 y_0$ and $x_2=\beta_2 y_{\tau}$. 

We next present a proof for the existence of the lower bound for $\eta^{(2)}$ for the TLS while operating as a heat engine the non-adiabatic regime $u \neq 1$.  

%\vspace{5mm}
\renewcommand{\theequation}{B\arabic{equation}}
\setcounter{equation}{0}  % reset counter

\section*{Appendix B: Proof for $\eta^{(2)} > \langle \eta \rangle^2$ for qubit Otto-engine}
To proceed with the proof in Eq.~(\ref{proof-TLS}), we simplify the following quantity
\bea
{\cal A} &&\equiv \langle w^2 \rangle_c \, \langle q_h \rangle^2 -\langle w \rangle^2 \, \langle q^2_h \rangle_c \nonumber \\
&& = u \, (u-1)\, \omega_0\, \omega_{\tau}^2 \, \Big[{\cal A}_1 + {\cal A}_2 \Big],
\eea
where 
%Since $  0 \leq  u \leq 1$, the first term in the above expression is negative, which imply 
\begin{widetext}
\bea
{\cal A}_1 &=& \Big[ 1 \!+ (1\!- 2u)\, {\bf t}(\bar{x}_0) \, {\bf t} (\bar{x}_{\tau}) \Big]\, \Big[ \omega_{\tau} {\bf t} (\bar{x}_0) \Big( (1\!-2u)\, {\bf t}(\bar{x}_0) \!+\! {\bf t} (\bar{x}_{\tau})\Big) + \omega_0 {\bf t}^2(\bar{x}_0)  \!+\! (1\!- 2 u) \omega_0 {\bf t}(\bar{x}_0) \, {\bf t}(\bar{x}_{\tau})  \Big]  \nonumber \\
{\cal A}_2 &=& \Big[(1\!- 2u) {\bf t}(\bar{x}_0) + {\bf t}(\bar{x}_{\tau}) \Big]\, \Big[\omega_{\tau} {\bf t}(\bar{x}_0) \big( 1-  {\bf t}^2(\bar{x}_{\tau})\big) - \omega_0\, {\bf t}(\bar{x}_{\tau}) \big[1 + (1\!- 2u) {\bf t}(\bar{x}_0) {\bf t}(\bar{x}_{\tau})\big] \Big]\nonumber
\label{A2}
\eea
where $\bar{x}_0= \beta_c \omega_0/2$ and $\bar{x}_{\tau}= \beta_h \omega_{\tau}/2$ and we have used a simplified notation for $\tanh$ function and write it as ${\bf t}$.
In the regime of heat engine operation, we demand $\langle q_h \rangle >0$ which implies the following condition
\be
(1- 2 u) {\bf t}(\bar{x}_0) + {\bf t}(\bar{x}_{\tau}) <0
\label{qh}
\ee
which also leads to a condition 
\be
1+ (1- 2 u) {\bf t}(\bar{x}_{0})\, {\bf t}(\bar{x}_{\tau}) < 1- {\bf t}^2(\bar{x}_{\tau}) 
\label{qh-1}
\ee
In addition to this, $\langle q_c \rangle = - \frac{\omega_{0}}{2} \Big[{\bf t}(\bar{x}_{0}) + {\bf t}(\bar{x}_{\tau}) (1-2u)\Big] <0$ implies,
\bea
1+ (1- 2 u) {\bf t}(\bar{x}_{0})\, {\bf t}(\bar{x}_{\tau}) > 1- {\bf t}^2(\bar{x}_{0}) >0 
\label{qc}
\eea
Furthermore, the requirement that the output work is positive i.e., $\langle w \rangle <0$ gives the  condition
\be
\big[ \omega_0 {\bf t}(\bar{x}_0) +\omega_{\tau} {\bf t}(\bar{x}_{\tau})\big] +(1-2 u)] \big[ \omega_0 {\bf t}(\bar{x}_{\tau}) + \omega_{\tau} {\bf t}(\bar{x}_0)\big] <0
\ee
Multiplying this with ${\bf t}(\bar{x}_0)$ and rearranging the terms, one receives,
\be
\Big[ \omega_{\tau} {\bf t} (\bar{x}_0) \Big( (1\!-2u)\, {\bf t}(\bar{x}_0) \!+\! {\bf t} (\bar{x}_{\tau})\Big) + \omega_0 {\bf t}^2(\bar{x}_0)  \!+\! (1\!- 2 u) \omega_0 {\bf t}(\bar{x}_0) \, {\bf t}(\bar{x}_{\tau})  \Big] <0
\label{work}
\ee
\end{widetext}
Similarly 
Combining Eq.~(\ref{qc})  and Eq.~(\ref{work}) confirms that ${\cal A}_1$ is always negative. Now the first part in ${\cal A}_2$ is negative which follows from Eq.~(\ref{qh}). Let us now look at the second term in ${\cal A}_2$, which using the condition Eq.~(\ref{qh-1}) provides a lower bound, given as
\bea
{\cal B} &=& \Big[\omega_{\tau} {\bf t}(\bar{x}_0) \big( 1\!-\! {\bf t}^2(\bar{x}_{\tau})\big) \!-\! \omega_0\, {\bf t}(\bar{x}_{\tau}) \big[1 \!+\! (1\!- 2u) {\bf t}(\bar{x}_0) {\bf t}(\bar{x}_{\tau})\big] \Big],\nonumber \\
{\cal B} &\geq& \Big[ \omega_{\tau} \, {\bf t}(\bar{x}_{0}) -  \omega_{0} \, {\bf t}(\bar{x}_{\tau})\Big] \, \Big[1 -  {\bf t}^2(\bar{x}_{\tau})\Big] 
\eea
while operating as an engine as $\frac{\omega_0}{\omega_{\tau}} \geq \frac {\beta_h}{\beta_c}$ which means $ {\bf t}(\bar{x}_{0})>  {\bf t}(\bar{x}_{\tau})$ and therefore ${\cal B} >0$. This completes our proof that $\eta^{(2)} > \langle \eta \rangle^2$ for a TLS working fluid working as an engine.
%In a simliar fashion, one can show that in the refrigerator regime,

\renewcommand{\theequation}{C\arabic{equation}}
\setcounter{equation}{0}  % reset counter

\section*{Appendix C: Characteristic function for harmonic oscillator system}
For the case when working fluid consisting of a simple harmonic oscillator, the exact expression for the CF was obtained earlier \cite{Denzler-thesis,Deffner-work} and is given as,
\begin{widetext}
\bea
\!\!\chi_{\rm HO} (\alpha_1,\alpha_2)&\!=\!& \frac{2}{{\cal Z}_0 {\cal Z}_{\tau}} \!\times \!\frac{1}{\sqrt{{\cal Q}\, (1\!-\! u_0^2) (1\!-\! v_0^2) +(1\!+ \! u_0^2) (1\!+\! v_0^2) \!-\!4 u_0 v_0}} \!\times\! \frac{1}{\sqrt{{\cal Q} \,(1\!-\! x_0^2) (1\!-\! y_0^2) +(1\!+\! x_0^2) (1\!+\! y_0^2) \!-\!4 x_0 y_0}},  \nonumber \\
\eea
\end{widetext}
where 
\bea
 u_0 &=& e^{-\omega_0(\beta_c + i \alpha_1)}, \nonumber \\
 v_0 &=& e^{i \omega_{\tau}(\alpha_1 - \alpha_2)}, \nonumber \\
  x_0 &=& e^{-\omega_{\tau}(\beta_h  + i (\alpha_1-\alpha_2))},\nonumber \\
   y_0 &=& e^{i \omega_0  \alpha_1},
\eea
and ${\cal Z}_0$ and ${\cal Z}_{\tau}$ are the partition functions. Here ${\cal Q} \in [1, \infty]$ is the so-called the adiabaticity parameter with the value 1 corresponding to the QS limit.  In this QS limit, the CF simplifies to,
\bea
\chi^{QS}_{\rm HO} (\alpha_1,\alpha_2)&=&   \frac{1}{{\cal Z}_0 {\cal Z}_{\tau}} \times \frac{1}{1-u_0 v_0} \times \frac{1}{1-x_0 y_0} 
\eea
%The marginal CF's for net work and absorbed heat are given as,
which can be further simplified and expressed in terms of the Bose-functions and given as,
\bea
\chi^{QS}_{\rm HO} (\alpha_1, \alpha_2)&&= \Big[1- n_{\tau} (1+n_0)\, \big(e^{-i\alpha_1(\omega_{\tau}-\omega_0)} \, e^{i \alpha_2\omega_{\tau}} -1 \big) \nonumber \\
&& - \, n_0(1+n_{\tau}) \, \big(e^{i\alpha_1 (\omega_{\tau}-\omega_0)} \, e^{-i\alpha_2 \omega_{\tau}} -1)\big)\Big]^{-1}, \nonumber \\ 
%\chi_{QS}^q(\alpha)&= &\Big[1-\big(n_{\tau}(1+n_0)(e^{i \alpha \omega_{\tau}}-1) \nonumber \\
%+&& n_0(1+n_{\tau})(e^{-i\alpha \omega_{\tau}}-1)\big)\Big]^{-1}
\eea
where $n_i= (e^{\beta_i \omega_i}-1)^{-1}$ is the Bose-Einstein distribution function. At this junction, it is interesting to compare the CF for HO and the TLS model which display differences between the two models in terms of the underlying statistics.   It is important to note that, the driven harmonic oscillator in the quasi-static (QS) limit, reduces to the hamiltonian of the form  $ H(t)=\omega(t)  a^{\dagger}a $  where $a^{\dagger}$ and $a$ are the creation and annihilation operators. 

\end{document}